\documentclass[sigconf,nonacm]{acmart}
\setcopyright{none}
\renewcommand\footnotetextcopyrightpermission[1]{}

\usepackage{listings}
\usepackage{float}
\usepackage{xcolor}
\usepackage[most]{tcolorbox}
\usepackage{capt-of}
\usepackage{graphicx}
\usepackage{tikz}
\usepackage{multirow}
\usepackage{svg}
\usepackage{algorithmic}
\usepackage{textcomp}
\usepackage{booktabs}
\usepackage{minted}
\usepackage{tabularx}

\usetikzlibrary{positioning,fit,arrows.meta,calc}

\usepackage[acronym]{glossaries}
\makeglossaries

\newacronym{mqss-db}{MQSS}{Munich Quantum Software Stack}

\newacronym{mqss}{MXS}{XXXXX Stack}

\newacronym{hpc}{HPC}{High Performance Computing}

\newacronym{lrz}{XXZ}{XX Supercomputing Centre}

\newacronym{hpcqc}{HPCQC}{High-Performance Computing-Quantum Computing}

\newacronym{qdmi}{QDMI}{Quantum Device Management Interface}

\newacronym{qpi}{QPI}{Quantum Programming Interface}

\newacronym{qc}{QC}{Quantum Computing}

\newacronym{dsl}{DSL}{Domain-Specific Language}

\newacronym{sdk}{SDK}{Software Development Kit}

\newacronym{gpu}{GPU}{Graphics Processing Unit}

\newacronym{fpga}{FPGA}{Field Programmable Gate Array}

\newacronym{qpu}{QPU}{Quantum Processing Unit}

\newacronym{qir}{QIR}{Quantum Intermediate Representation}

\newacronym{ir}{IR}{Intermediate Representation}

\newacronym{qrmi}{QRMI}{Quantum Resource Management Interface}

\newacronym{jit}{JIT}{Just-in-Time}

\newacronym{cc}{CC}{Client Creation}

\newacronym{rq}{RQ}{Resource Query}

\newacronym{jsc}{JSC}{Job Submission and Creation}

\newacronym{eqc}{EQC}{Execution of Quantum Circuit}

\newacronym{res}{RES}{Result Retrieval}

\newacronym{te}{TE}{Total Execution}

\newacronym{iqr}{IQR}{Interquartile range}

\newacronym{vqe}{VQE}{Variational Quantum Eigensolver}

\newacronym{qft}{QFT}{Quantum Fourier Transform}

\newacronym{mlir}{MLIR}{Multi-Level Intermediate Representation}

\newacronym{api}{API}{Application Programming Interface}

\newacronym{saxpy}{SAXPY}{Single-Precision A*X Plus Y}

\newacronym{qpir}{QPIR}{Quantum Programming Intermediate Representation}

\usepackage[T1]{fontenc}
\usepackage[utf8]{inputenc}

\setminted{fontsize=\small}
\setminted{breaklines}

\lstdefinestyle{cppstyle}{
    language=C++,
    basicstyle=\ttfamily\small,
    numbers=left,
    numberstyle=\tiny,
    stepnumber=1,
    numbersep=8pt,
    xleftmargin=20pt,
    showstringspaces=false,
    breaklines=true,
    columns=fullflexible,
    keywordstyle=\color{blue},
    commentstyle=\color{green!50!black},
    stringstyle=\color{red!70!black},
    frame=none
}

\begin{document}

\raggedbottom
\title{OpenMQPI: An MLIR-based Pipeline to Offload Quantum Circuits via OpenMP}

\author{Ercüment Kaya}
\authornote{Both authors contributed equally to this research.}
\email{ercument.kaya@tum.de}
\affiliation{%
  \institution{Leibniz Supercomputing Centre}
  \institution{Technical University of Munich}
  \city{Garhing near Munich}
  \country{Germany}}

\author{Ekin Devim Şahinkaya}
\email{ekin.sahinkaya@tum.de}
\authornotemark[1]
\affiliation{%
  \institution{Leibniz Supercomputing Centre}
  \institution{Technical University of Munich}
  \city{Garhing near Munich}
  \country{Germany}}

\author{Minh Chung}
\email{minh.chung@lrz.de}
\affiliation{%
  \institution{Leibniz Supercomputing Centre}
  \city{Garhing near Munich}
  \country{Germany}}

\author{Martin Schulz}
\email{schulzm@in.tum.de}
\affiliation{%
  \institution{Technical University of Munich}
  \institution{Leibniz Supercomputing Centre}
  \city{Garhing near Munich}
  \country{Germany}
}

\begin{abstract}
The emerging field of quantum computing requires specialized software stacks and programming interfaces.
However, these tools are often tightly coupled to specific software ecosystems and hardware backends, limiting interoperability and software reuse. 
For \gls{hpcqc} application developers, this complicates the integration of quantum accelerators into existing applications.

Existing quantum programming frameworks expose circuit representations and compilation pipelines that are tightly coupled to their execution ecosystems, making it difficult to integrate quantum compilation with established heterogeneous compiler infrastructures.
To address this issue, we propose OpenMQPI, a \gls{mlir}-based OpenMP extension for quantum programming.
OpenMQPI leverages OpenMP's standardized target-offloading infrastructure to provide a reusable approach for integrating quantum resources into \gls{hpc} applications.
To enable quantum offloading with OpenMQPI, we create QPIR, a lightweight \gls{mlir} dialect for representing quantum circuits inside the  OpenMP target regions, together with a compiler pipeline that performs quantum-specific optimizations and lowers the resulting representation into an executable compatible with the OpenMP offloading model.
At runtime, the generated executable leverages the OpenMP target offloading model via a MQSS plugin within OpenMQPI to offload quantum circuits to the quantum hardware.

We evaluate OpenMQPI using the \gls{vqe}, the \gls{qft}, and the Deutsch--Jozsa algorithm with 5, 10, 15, and 20 qubits, and compare its runtime performance against Qiskit and CUDA Quantum.
The results show that OpenMQPI achieves the lowest average execution time for \gls{vqe} across all evaluated qubit configurations.
At 20 qubits, OpenMQPI requires 5.18~s, compared with 5.83~s for Qiskit and 7.39~s for CUDA Quantum.
For \gls{qft} and Deutsch--Jozsa, OpenMQPI remains close to Qiskit while achieving lower execution times than CUDA Quantum.

\end{abstract}

%



\maketitle

\section{Introduction}

Quantum computing has the potential to address computational problems that are believed to be unsolvable for classical computers~\citep{preskill2018quantum}.
In practice, however, many quantum applications also rely on classical computation to support and coordinate quantum execution~\citep{callison2022hybrid, 10821378, 10.1145/3637543.3653432}.
This has motivated growing interest in integrating quantum computing resources into \gls{hpc} environments, where classical and quantum resources can work together within hybrid applications~\citep{beck2024integrating}.

Conspicuous progress has been made in the development of quantum computing hardware \citep{mansfield2025iqmfirstpractice, neven-2025}. 
However, to achieve the full potential of \gls{qc}, this hardware requires compatible programming languages, interfaces, and compilation infrastructure that can coordinate classical and quantum resources within a common execution environment.


A variety of programming interfaces, including Qiskit\citep{qiskit}, PennyLane\citep{pennylane}, and CUDA Quantum\citep{cuda-q}, provide tools for developing and executing quantum applications.
However, these frameworks rely on different programming models, compiler infrastructures, and execution environments.
As a result, integrating quantum resources into existing \gls{hpc} applications often requires developers to work with framework-specific software stacks and interfaces.
This framework-specific approach limits the portability and reusability of software components across different \gls{hpcqc} environments.

A key challenge is that traditional \gls{hpc} offloading and quantum execution expose different compilation and execution models.
OpenMP target offloading is designed around launching compiled device code on a target device, whereas a quantum resource executes a circuit or quantum job rather than conventional device instructions.

Therefore, the quantum circuit must remain available as an explicit representation beyond the device-code lowering stage. 
Directly treating a quantum resource as another OpenMP target device does not provide a natural mechanism to preserve and optimize the circuit while still using the OpenMP compilation and runtime infrastructure.

This observation motivates a different approach: treating quantum circuits as first-class target payloads rather than as machine code. 
OpenMQPI preserves the circuit at the compiler level, applies quantum-specific transformations in \gls{mlir}, and then transports the resulting circuit through the OpenMP target-offloading path to a quantum-specific runtime plugin.

In this paper, we introduce OpenMQPI, an \gls{mlir}-based compilation infrastructure that integrates quantum circuit compilation and execution into the OpenMP target-offloading model.
Rather than introducing another framework-specific programming interface, OpenMQPI builds on the existing OpenMP infrastructure, allowing quantum resources to be integrated as target devices within the programming model already widely used in HPC.

OpenMQPI preserves the quantum circuit as an \gls{ir} during the compilation rather than lowering it to target device instructions.
The \gls{mlir}-based representation enables quantum-specific optimizations before the circuits are lowered for execution.
The resulting circuit is then carried through the OpenMP offloading pipeline and delivered to a quantum-specific \texttt{libomptarget} plugin, which submits the circuit to the selected quantum resource.


The implementation leverages the OpenMP runtime and MQSS Client interface \citep{kaya2026mqssclientinterfacedecoupling} to submit quantum circuits to the selected quantum device. The approach is considered a custom \texttt{libomptarget} plugin, referred to as the MQSS plugin.
%
We evaluate OpenMQPI using the \gls{vqe}, the \gls{qft}, and the Deutsch--Jozsa algorithm across different qubit configurations.
Our evaluation validates the complete compilation-and-execution pipeline and demonstrates that OpenMQPI maintains competitive runtime performance compared with existing quantum programming approaches.

This paper makes the following contributions:
\begin{itemize}
    \item \textbf{An \gls{mlir}-based representation for quantum circuits:} We introduce \gls{qpir}, a lightweight \gls{mlir} dialect for representing and optimizing quantum circuits.

    \item \textbf{Circuit-preserving integration with OpenMP target offloading}: We extend the OpenMP target-offloading infrastructure to represent quantum circuits as target payloads, preserve them through compilation, and deliver them to the MQSS plugin for execution.

    \item \textbf{A compiler and runtime pipeline for quantum execution:}We implement qpir\_cc, qpir\_opt, and qpir\_link together with an MQSS plugin to generate, optimize, lower, and execute quantum circuits through the OpenMP offloading infrastructure.

    \item \textbf{An experimental evaluation of OpenMQPI:} We evaluate the proposed infrastructure and its runtime performance across different quantum applications.
\end{itemize}

The remainder of this paper is organized as follows: Section~\ref{sec:background} covers background.
In Section~\ref{sec:openmqpi}, we introduce the OpenMQPI, discuss its architecture and components.
In Section~\ref{sec:evaluation}, we present the experiments and performance evaluation.
Section ~\ref{sec:related-works} reviews the related literature.
Finally, we conclude the paper and discuss future work in Section~\ref{sec:conclusion}.
\section{Background}
\label{sec:background}

This section briefly explains the basics of the concepts and tools that we exploit to keep this paper self-contained.

\subsection{\gls{hpc} and \gls{qc}}

\gls{hpc} uses supercomputers and computer clusters to address complex computational problems with large data sets, large-scale simulations, and the modeling of complex physical phenomena~\citep{holmes2016hpc, vetter2013contemporary}.

The \gls{hpc} applications often use parallel processing and computation offloading to hardware accelerators such as \gls{gpu}, \gls{fpga}, and quantum computers to improve the performance of \gls{hpc} applications~\citep{aamodt2018general, vanderbauwhede2013high, nielsen2000quantum}.
The many-core architecture of \gls{gpu}s is highly suitable for data-parallel processing, while quantum computers use a fundamentally different computational model.
%
Quantum computers use radically different modalities that could change the complexity of several computational problems, from factorization to optimization~\citep{kaye2006introduction}.
However, this potential power comes with new requirements on users, \gls{hpc} centers, and the developers.

The requirements include, but are not limited to, programming interfaces, compilation infrastructure, and a hybrid \gls{hpcqc} laboratory~\citep{10.1145/3637543.3653432}.
%
Programming interfaces, such as Qiskit~\citep{qiskit} and CUDA-Q~\citep{cuda-q}, are used to describe quantum circuits in \gls{hpcqc} applications.
A compilation infrastructure that is tailored for \gls{qc} is required to optimize and ensure that the given circuit is executable on the target hardware.

\subsection{LLVM and \gls{mlir}}

LLVM is a compiler framework that can be used to develop front-ends for programming languages and back-ends for architectures \citep{lattner2004llvm}.
The LLVM \citep{llvm} project contains multiple sub-projects, including compiler front-ends such as Clang \citep{clang}, and Flang \citep{flang} to a debugger named LLDB\citep{lldb}. 

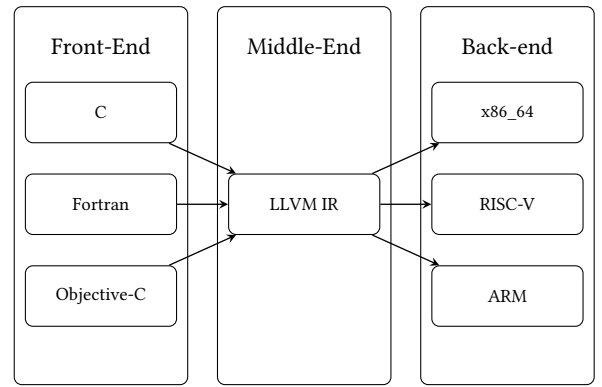
\begin{figure}[ht]
    \centering
    \begin{tikzpicture}[
        node distance=0.38cm,
        >=stealth,
        container/.style={
            draw,
            rounded corners=1mm,
            text width=0.25\linewidth,
            minimum height=5.0cm,
            inner sep=2.5pt,
            font=\footnotesize
        },
        subcontainer/.style={
            draw,
            rounded corners=1mm,
            minimum width=2cm,
            minimum height=0.8cm,
            align=center,
            font=\footnotesize
        },
        arrow/.style={
            ->,
            line width=0.5pt
        }
    ]

    \node[container, anchor=north east] (front) {};
    \node[container, right=of front] (middle) {};
    \node[container, right=of middle] (back) {};

    \node[anchor=north] at ([yshift=-3mm]front.north) {Front-End};
    \node[anchor=north] at ([yshift=-3mm]middle.north) {Middle-End};
    \node[anchor=north] at ([yshift=-3mm]back.north) {Back-end};

    \node[subcontainer, anchor=north] (f1)
        at ([yshift=-10mm]front.north) {C};

    \node[subcontainer, below=0.4cm of f1] (f2)
        {Fortran};

    \node[subcontainer, below=0.4cm of f2] (f3)
        {Objective-C};

    \node[subcontainer] (m1)
        at (middle.center |- f2.center) {LLVM IR};

    \node[subcontainer, anchor=north] (b1)
        at ([yshift=-10mm]back.north) {x86\_64};

    \node[subcontainer, below=0.4cm of b1] (b2)
        {RISC-V};

    \node[subcontainer, below=0.4cm of b2] (b3)
        {ARM};
        
    \draw[arrow] (f1) -- (m1);
    \draw[arrow] (f2) -- (m1);
    \draw[arrow] (f3) -- (m1);
    
    \draw[arrow] (m1) -- (b1);
    \draw[arrow] (m1) -- (b2);
    \draw[arrow] (m1) -- (b3);

    \end{tikzpicture}

    \caption{Simplified Version of LLVM's Workflow}
    \label{fig:llvm}
\end{figure}

As illustrated in Figure \ref{fig:llvm}, LLVM follows a modular approach; therefore, high-level programming languages, such as C and Fortran, are lowered to common LLVM \gls{ir} using Clang or Flang.
The generated \gls{ir} can then be optimized using LLVM passes before that, being lowered to machine code for the target architecture.
This approach enables LLVM to provide language- and target-independent optimizations, while also supporting target-specific optimizations.

LLVM \gls{ir} is designed to resemble machine code, enabling domain-agnostic optimizations.
However, this approach loses explicit domain information, such as  matrix multiplication and hardware accelerators. 
To address this limitation, Lattner et al. propose \gls{mlir} \citep{mlir-1, mlir-2}. 
It provides a higher-level abstraction that preserves explicit information and progressively lowers it to LLVM \gls{ir} to enable better optimizations.

The \gls{mlir} infrastructure provides various, extensible dialects to distinguish domains.
An \gls{mlir} dialect includes, but is not limited to, domain-specific operations and types.
Hence, optimizers can work on the different parts of the code that make sense for them. 

\subsection{OpenMP}

OpenMP \citep{dagum1998openmp} is an \gls{api} for shared-memory parallel programming in C, C++, and Fortran, and is widely used in the \gls{hpc} domain to parallelize the application with a set of compiler directives and runtime library routines.
Thanks to this approach, developers can parallelize \gls{hpc} applications without major changes to the source code.
OpenMP follows the fork-join execution model, that is, an application starts with a single thread, and after reaching a parallel region, additional threads are created to execute the region in parallel.
Once the execution of the parallelized section is completed, the threads join the original thread and continue the execution of the application.

OpenMP was initially designed for parallel programming on CPUs.
However, with the 4.0 specification \citep{openmp40}, OpenMP introduces support for target offloading to enable the use of hardware accelerators such as \gls{gpu}.
This model allows developers to use a single source code for different architectures while improving portability across heterogeneous computing systems \citep{bertolli2015integrating}.

Listing \ref{lst:openmp_offload} contains an \gls{saxpy} routine to be offloaded to the target device.
The keyword \texttt{target} specifies that the region should be executed on a target device.
\texttt{target data} creates a region where data can be mapped to the device.
\texttt{map} controls how variables are transferred between the host and the device.
\texttt{teams} creates groups of threads on the devices, while \texttt{distribute} partitions the iterations among the created teams.
\texttt{parallel} enables multiple threads in each team to execute simultaneously.

\definecolor{qpirbackground}{RGB}{248,248,248}
\definecolor{qpirborder}{RGB}{190,190,190}

\lstdefinestyle{qpirstyle}{
    basicstyle=\ttfamily\scriptsize,
    columns=fullflexible,
    keepspaces=true,
    showstringspaces=false,
    breaklines=true,
    frame=none,
    aboveskip=0pt,
    belowskip=0pt
}


\begin{listing}[ht]
\begin{lstlisting}[style=cppstyle]
void saxpy() {
float a, x[N], y[N];
...
#pragma omp target data map(to: x[0:N], a) map(tofrom: y[0:N])
{
    #pragma omp target teams distribute parallel for
    for (int i = 0; i < N; i++) {
        y[i] = a * x[i] + y[i];
    }
}

\end{lstlisting}
\caption{An example of \gls{saxpy} routine, implemented in OpenMP}
\label{lst:openmp_offload}
\end{listing}


The OpenMP offloading mechanism requires compiler and runtime support.
While the compiler generates code for the host and the target device, OpenMP offloading runtime libraries, such as \texttt{libomptarget} in LLVM and \texttt{libgomp} in GCC, manage the execution of regions on the target device.
In LLVM, \texttt{libomptarget} provides a generic runtime for managing OpenMP offloading and communicates with device-specific implementations through a plugin interface. 
These plugins handle hardware-specific operations, such as device initialization, memory allocation, data transfers between host and device, kernel execution, and synchronization.
This separation allows the generic OpenMP runtime to support different accelerator architectures while delegating device-specific functionality to the corresponding plugin.

\section{OpenMQPI}
\label{sec:openmqpi}
\subsection{Motivation}
OpenMQPI is motivated by the idea of extending OpenMP's established target-offloading model beyond classical accelerators to support quantum computing resources.
This allows quantum resources to be integrated into an existing \gls{hpc} programming and runtime infrastructure rather than requiring a separate execution model.
At the compiler level, \gls{mlir} is particularly well-suited to this approach because its dialect-based design enables domain-specific representations to be introduced and optimized while remaining within a common compiler infrastructure.
OpenMQPI leverages this flexibility to represent quantum operations through the \gls{qpir} dialect and progressively lower them into the OpenMP offloading pipeline.
\subsection{Design and Components}
OpenMQPI consists of a set of compiler and runtime components that extend the OpenMP offloading workflow for quantum circuit execution.
On the compiler side, the \gls{qpir} dialect provides the intermediate representation for quantum circuits, while \texttt{qpir\_cc}, \texttt{qpir\_opt}, and \texttt{qpir\_link} handle circuit generation, optimization, and lowering.
On the runtime side, the MQSS plugin extends \texttt{libomptarget} to execute the resulting quantum circuits on quantum resources through the MQSS Client interface.

\subsubsection{\gls{qpir} Dialect}
QPIR is a lightweight \gls{mlir} dialect designed to represent quantum circuits within the OpenMQPI compilation pipeline.
During its design, we considered existing quantum representations, including CUDA Quantum's \citep{cuda-q} Quake \citep{nvidia_quake_dialect} and Catalyst's \citep{ittah2024catalyst} Quantum \citep{pennylane_catalyst_quantum_dialect} dialects.
However, adopting these dialects would introduce additional dependencies on their corresponding software ecosystems.

The \gls{qpir} dialect follows an imperative programming model and uses an identifier-based representation of quantum resources, where qubits are addressed by integer identifiers rather than represented as first-class SSA values.

\definecolor{qpirbackground}{RGB}{248,248,248}
\definecolor{qpirborder}{RGB}{190,190,190}

\lstdefinestyle{qpirstyle}{
    basicstyle=\ttfamily\scriptsize,
    columns=fullflexible,
    keepspaces=true,
    showstringspaces=false,
    breaklines=true,
    frame=none,
    aboveskip=0pt,
    belowskip=0pt
}



\begin{listing}[ht]
\begin{lstlisting}[style=qpirstyle]
omp.target map_entries(%2 -> %arg3, %4 -> %arg4, %5 -> %arg5 :
    !llvm.ptr, !llvm.ptr, !llvm.ptr) {
  %6 = llvm.load %arg3 : !llvm.ptr -> f64
  %7 = llvm.load %arg4 : !llvm.ptr -> f64
  %c0_i32 = arith.constant 0 : i32
  %c1_i32 = arith.constant 1 : i32
  qpir.h %c0_i32 : i32
  qpir.rx(%6) %c0_i32 : f64, i32
  qpir.cx %c0_i32, %c1_i32 : i32, i32
  qpir.ry(%7) %c1_i32 : f64, i32
  qpir.barrier
  %8 = qpir.measure %c0_i32 : i32 -> i1
  %9 = qpir.measure %c1_i32 : i32 -> i1
  omp.terminator
}
\end{lstlisting}
\caption{Example of the QPIR dialect within an OpenMP target region.}
\label{lst:qpir-example}
\end{listing}


Listing~\ref{lst:qpir-example} shows an example \gls{qpir} circuit represented within an OpenMP target region.
Quantum gates are represented as explicit \gls{qpir} operations, such as \texttt{qpir.h}, \texttt{qpir.rx}, \texttt{qpir.cx}, and \texttt{qpir.ry}, while measurement and synchronization are represented by \texttt{qpir.measure} and \texttt{qpir.barrier}.
Qubits are identified using integer indices, while rotation angles are represented as floating-point values.
By keeping these operations within \texttt{omp.target}, \gls{qpir} represents the quantum circuit while preserving the OpenMP target structure required by the subsequent offloading stages.
\subsubsection{QPIR\_CC}

The \texttt{qpir\_cc} component acts as the compiler front end of OpenMQPI and generates the quantum circuit as QPIR operations within the device-side MLIR.

\begin{figure}[!ht]
    \centering    \includegraphics[width=\columnwidth]{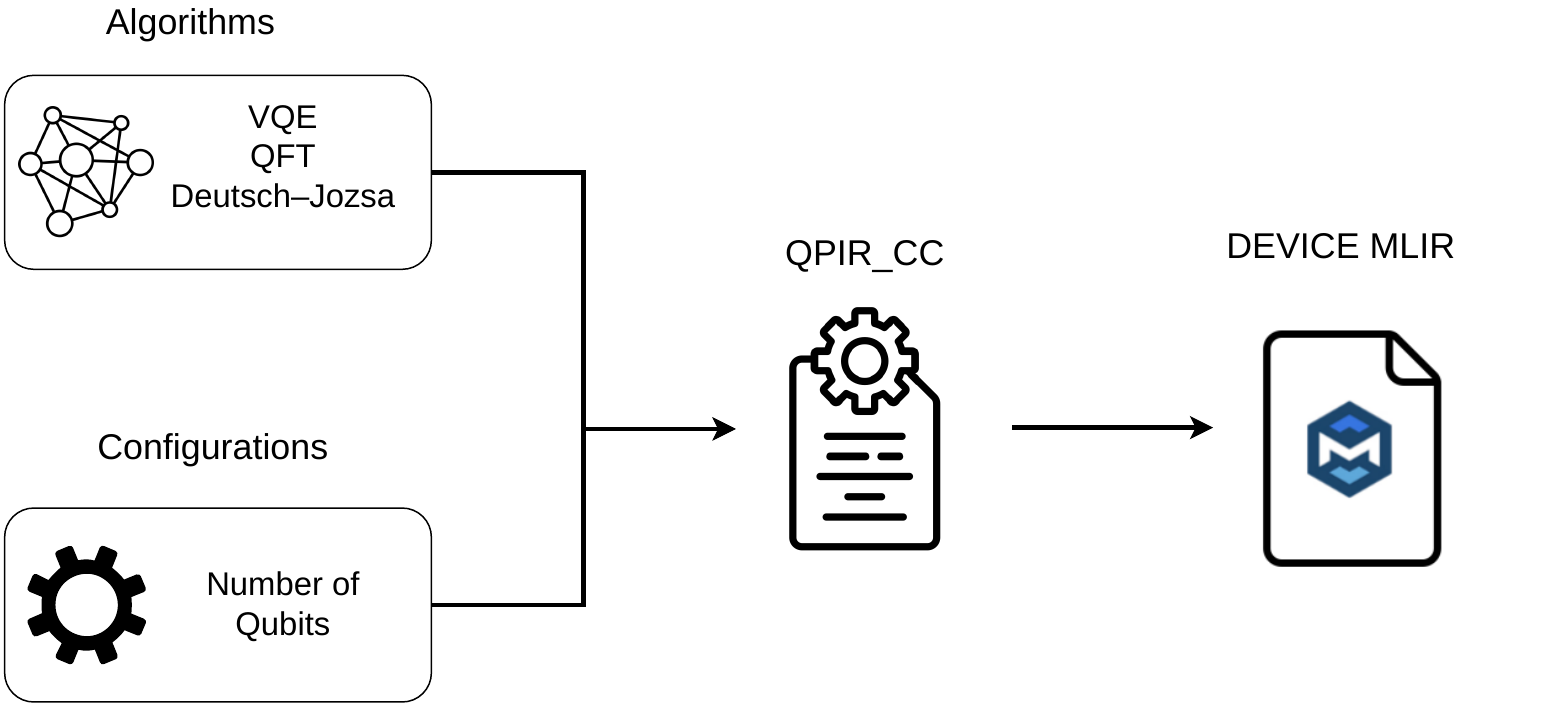}
    \caption{Generation of an \gls{mlir} file from algorithm and configuration inputs using \texttt{qpir\_cc}.}
    \label{fig:qpir-cc-workflow}
\end{figure}

It is able to create quantum circuits for \gls{vqe}, \gls{qft}, and Deutsch–Jozsa algorithms with multiple qubit configurations as illustrated in Fig.~\ref{fig:qpir-cc-workflow}.
While the inputs of the \texttt{qpir\_cc} are the algorithm as a string and the number of qubits as an integer, the output is the corresponding quantum circuit as an \gls{mlir} file.

As shown in Listing~\ref{lst:qpir-cc}, the generated device code that corresponds to the Bell State is placed inside the \texttt{omp.target} region, which allows for following the OpenMP offloading path.
The module attributes, such as \texttt{omp.is\_target\_device} and \texttt{omp.is\_gpu} direct the target region to the MQSS plugin.
%
Therefore, the resulting \gls{mlir} file contains the quantum circuit representation as well as the OpenMP information required by the subsequent optimization and lowering stages.
\definecolor{qpirbackground}{RGB}{248,248,248}
\definecolor{qpirborder}{RGB}{190,190,190}

\lstdefinestyle{qpirstyle}{
    basicstyle=\ttfamily\scriptsize,
    columns=fullflexible,
    keepspaces=true,
    showstringspaces=false,
    breaklines=true,
    frame=none,
    aboveskip=0pt,
    belowskip=0pt
}


\begin{listing}[ht]
\begin{lstlisting}[style=qpirstyle]
module attributes {llvm.data_layout = "e-p:64:64-i64:64-n8:16:32:64", llvm.target_triple = "qpu", omp.is_gpu = false, omp.is_target_device = true, omp.requires = #omp<clause_requires none>, omp.version = #omp.version<version = 11>, qpir.circuit_format = "qpir", qpir.target = "mqss_plugin"} {
  func.func @bell_state(%arg0: !llvm.ptr) {
    %0 = llvm.mlir.constant(1 : i64) : i64
    %1 = omp.map.info var_ptr(%arg0 : !llvm.ptr, i8) map_clauses(from) capture(ByRef) -> !llvm.ptr {name = "qpi_result"}
    omp.target map_entries(%1 -> %arg1 : !llvm.ptr) {
      %c0_i32 = arith.constant 0 : i32
      %c1_i32 = arith.constant 1 : i32
      qpir.h %c0_i32 : i32
      qpir.cx %c0_i32, %c1_i32 : i32, i32
      qpir.barrier
      %2 = qpir.measure %c0_i32 : i32 -> i1
      %3 = qpir.measure %c1_i32 : i32 -> i1
      omp.terminator
    }
    return
  }
}

}
\end{lstlisting}
\caption{The Device Side Code, generated by the QPIR\_CC}
\label{lst:qpir-cc}
\end{listing}


%
%
%

\subsubsection{QPIR\_OPT}
\begin{figure}[!ht]
    \centering
    \includegraphics[width=0.55\columnwidth]{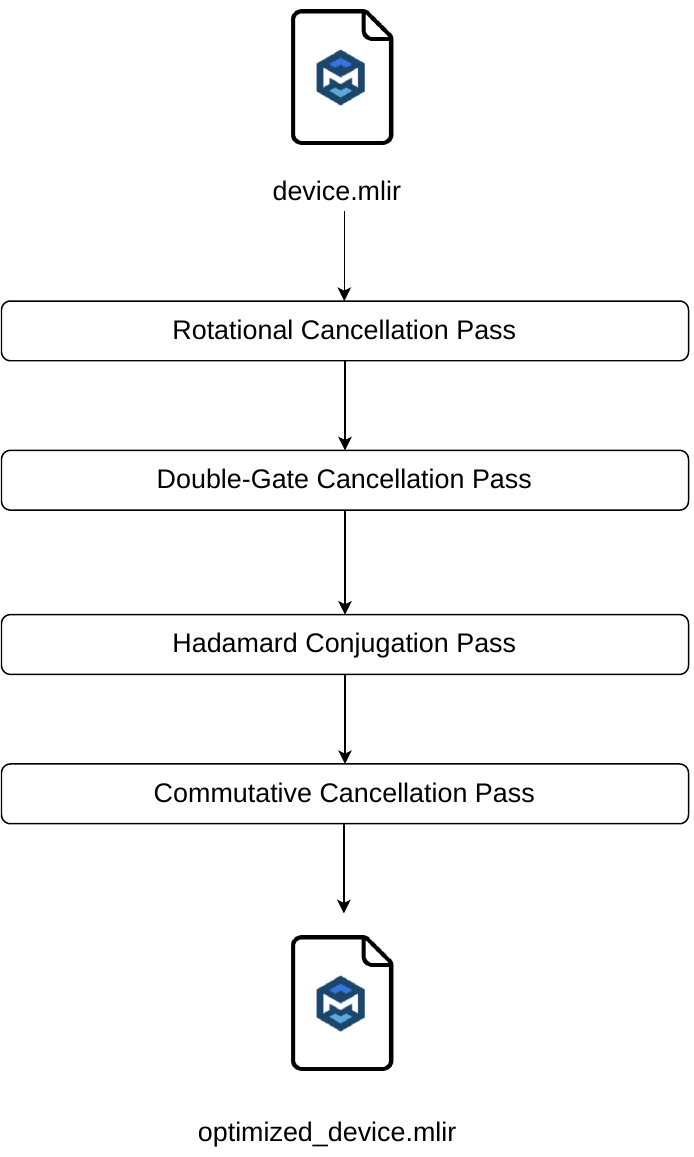}
    \caption{\gls{qpir}-specific optimization passes applied by
    \texttt{qpir\_opt}.}
    \label{fig:qpir-opt-passes}
\end{figure}

The \texttt{qpir\_opt} component performs optimizations specific to
quantum circuits on the \gls{mlir} module generated by
\texttt{qpir\_cc}.
As shown in Fig.~\ref{fig:qpir-opt-passes}, it simplifies and optimizes the quantum circuits by executing the following optimization passes on the \gls{qpir} dialect, respectively: 
\begin{itemize}
    \item \textbf{Rotational Cancellation:} This pass simplifies the quantum circuit by summing and normalizing the angles of rotation gates applied sequentially to the same qubit.
    \item \textbf{Double-Gate Cancellation:} This pass cancels the sequentially applied self-inverse gates, such as CNOT and Hadamard.
    \item \textbf{Hadamard Conjugation:} This pass reduces the circuit depth by replacing H-X-H and H-Z-H gate sequences with Z and X gates, respectively.
    \item \textbf{Commutative Cancellation:} This pass simplifies controlled-gate sequences by replacing H-CX-H and H-CZ-H patterns on the target qubit with CZ and CX gates, respectively.
\end{itemize}
Each transformation preserves the semantics of the original circuit while reducing the depth of the quantum circuit to be executed.
%

%

\subsubsection{QPIR\_Link}

\begin{figure}[!ht]
    \centering
    \includegraphics[width=\columnwidth]{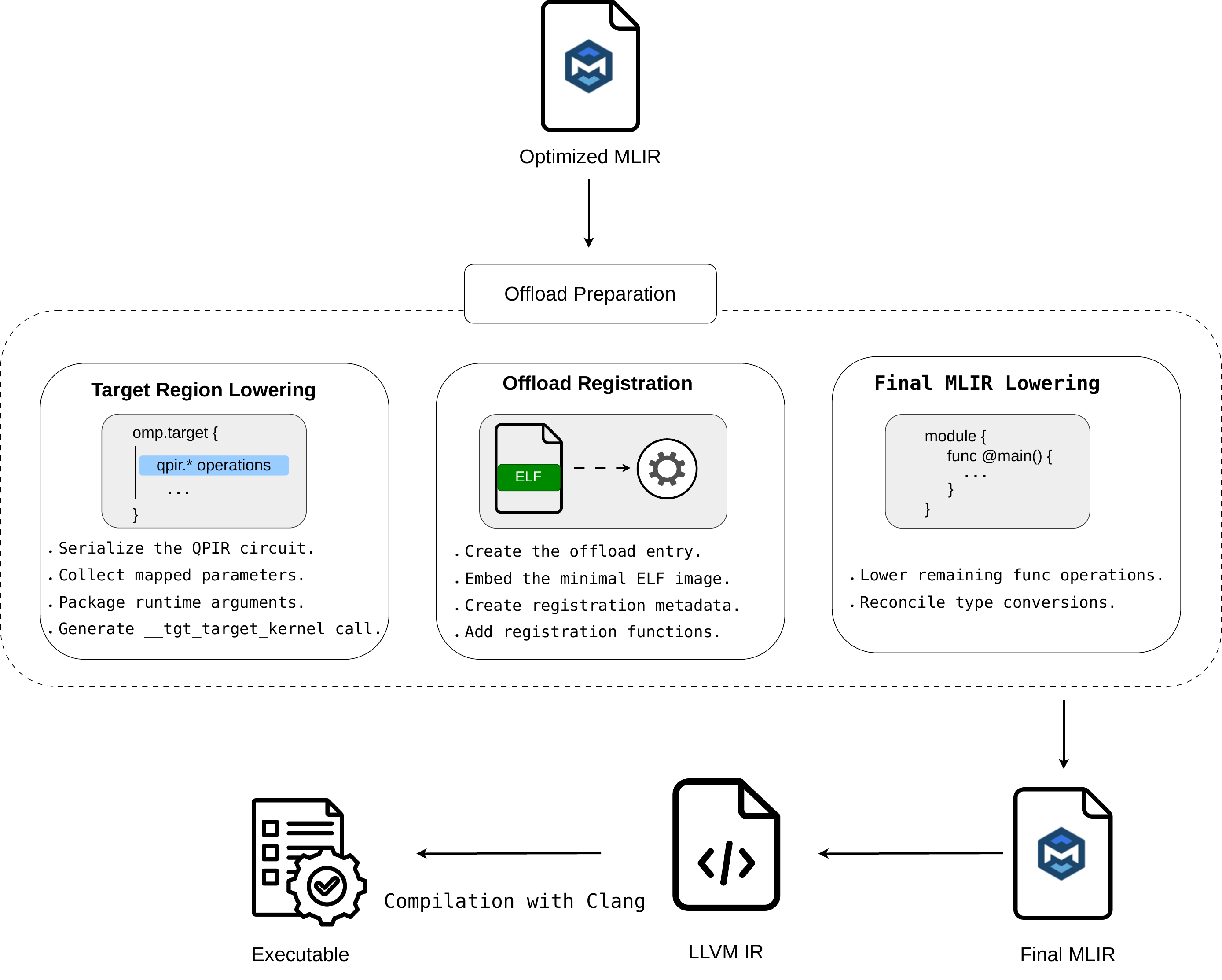}
    \caption{Overview of the \texttt{qpir\_link} lowering and executable generation workflow.}
    \label{fig:qpir-link}
\end{figure}

The \texttt{qpir\_link} is the final component of the OpenMQPI's compilation pipeline.
As illustrated in Fig.~\ref{fig:qpir-link}, it serializes the quantum circuit, performs offload registration, lowers the remaining module to LLVM \gls{ir}, and links it with the host code to create an executable.

%
%

During target-region lowering, the quantum circuit is serialized into a textual representation rather than being converted into target-specific machine instructions.
Runtime parameters are collected from the OpenMP mapping information and packaged together with the circuit into the data structures expected by \texttt{libomptarget}.
The \texttt{omp.target} operation is then replaced with a call to \texttt{\_\_tgt\_target\_kernel}, enabling the circuit and its runtime parameters to follow the standard OpenMP offloading path.

The offload-registration stage adds the metadata required by \texttt{libomptarget}, including an offload entry, a device-image descriptor, and the corresponding registration functions.
Since OpenMQPI transfers the quantum circuit as a runtime argument rather than executable device code, \texttt{qpir\_link} embeds a minimal ELF image to satisfy the device-image structure expected by the OpenMP runtime.
The MQSS plugin does not execute this image; the quantum circuit is provided separately through the runtime arguments.

After the target region and offload metadata have been prepared, the remaining \gls{mlir} operations are lowered to the LLVM dialect, and intermediate type conversions are resolved.
The resulting \gls{mlir} module is translated into LLVM IR and compiled with Clang to generate the final executable.

Crucial aspect of this lowering is to preserve the quantum circuit as a compiler-level payload rather than lowered into device instructions.
Therefore, the OpenMP runtime provides the offloading mechanism, while the quantum-specific plugin provides the execution mechanism.
This separation allows the existing OpenMP target-offloading infrastructure to be reused without requiring the quantum resource to execute an OpenMP-generated device binary.

\subsubsection{MQSS Plugin}
\begin{figure}[!ht]
    \centering
    \includegraphics[width=\columnwidth]{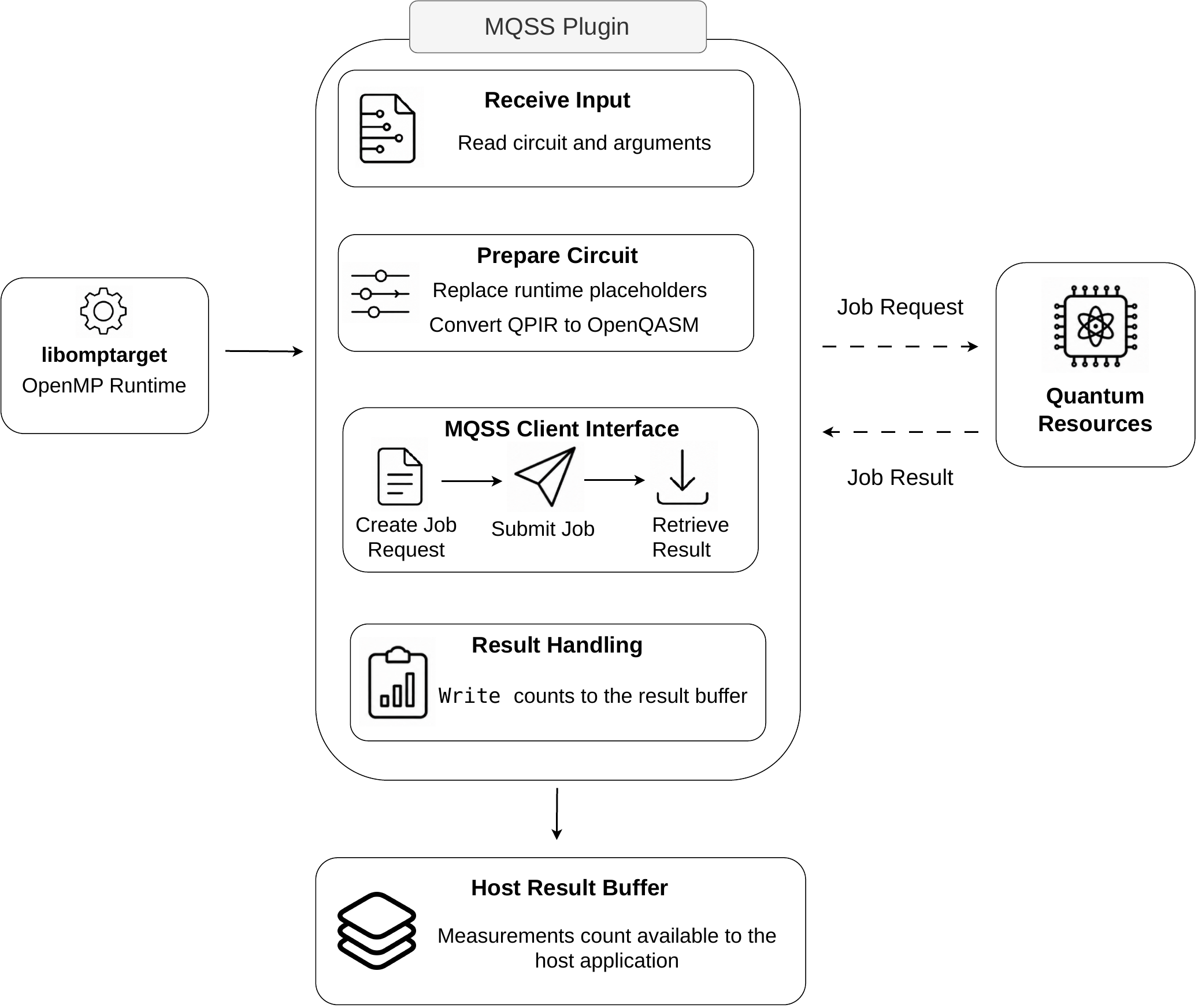}
    \caption{Overview of the MQSS plugin and its runtime execution workflow.}
    \label{fig:mqss-plugin}
\end{figure}

The MQSS plugin extends the \texttt{libomp\-target}  plugin infrastructure to support quantum resources as OpenMP target devices.
As illustrated in Fig.~\ref{fig:mqss-plugin}, \texttt{libomp\-target} forwards the circuit and its runtime arguments to the plugin when the target region is launched.
The plugin then prepares the circuit for execution and uses the MQSS Client interface to communicate with the selected quantum resource.

The plugin first reads the textual circuit representation and the associated runtime arguments.
Runtime-dependent values, such as rotation angles, are inserted into the circuit by replacing the placeholders generated during lowering.
The circuit is then converted from \gls{qpir} to OpenQASM before being submitted for execution.

The prepared circuit is submitted to the selected quantum resource as a job through the MQSS Client interface.
%
After execution, the plugin retrieves the measurement results and writes the returned counts into the result buffer provided by the host application.
The results, therefore, become available to the host through the same OpenMP offloading execution path.
\subsection{Compilation and Runtime Workflow}

\begin{figure*}[t]
    \centering
    \includegraphics[width=0.85\textwidth]{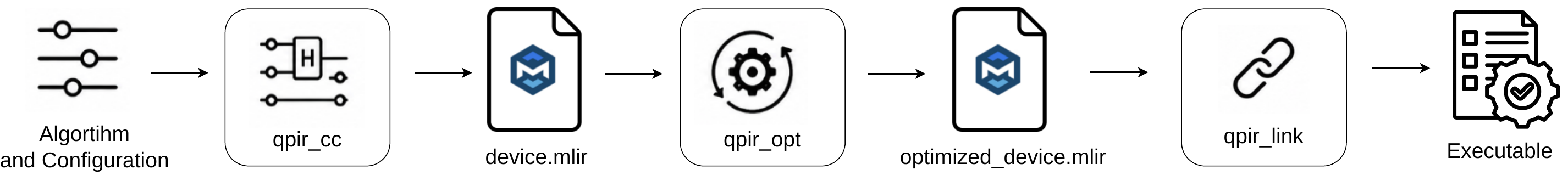}
    \caption{Overview of the OpenMQPI compilation workflow.}
    \label{fig:compilation-workflow}
\end{figure*}

Fig.~\ref{fig:compilation-workflow} summarizes the end-to-end compilation workflow of OpenMQPI.
Starting from an algorithm and its configuration, \texttt{qpir\_cc} generates an \gls{mlir} file containing the quantum circuit represented with \gls{qpir} operations.
The generated \gls{mlir} is then processed by \texttt{qpir\_opt}, which applies the quantum-specific optimization passes described above.
Finally, \texttt{qpir\_link} prepares the optimized \gls{mlir} for OpenMP target offloading and lowers it toward LLVM IR, from which the final executable is generated.

\begin{figure}[H]
    \centering    \includegraphics[width=0.95\columnwidth]{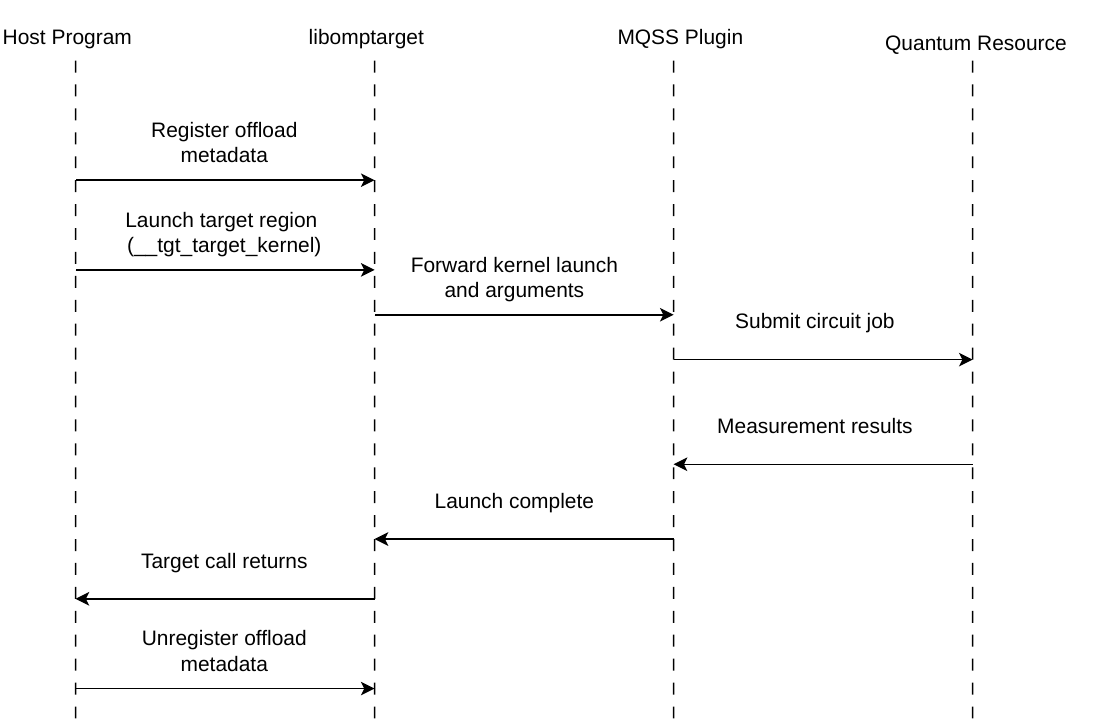}
    \caption{Runtime execution workflow of OpenMQPI through the OpenMP offloading infrastructure.}
    \label{fig:runtime-workflow}
\end{figure}

The runtime workflow is illustrated in Fig.~\ref{fig:runtime-workflow}.
When the executable starts, the generated offload metadata is registered with \texttt{libomptarget}.
Upon reaching the OpenMP target region, the host program invokes \texttt{\_\_tgt\_target\_kernel}, and \texttt{libomptarget} forwards the kernel launch and its arguments to the MQSS plugin.
The plugin submits the prepared circuit to the selected quantum resource and waits for results.
Once the measurement results are returned, the plugin completes the offload operation, and control returns to the host application through \texttt{libomptarget}.
Finally, the registered offload metadata is released when the application terminates.

\section{Evaluation}
\label{sec:evaluation}
This evaluation addresses two questions.
\textbf{RQ1: End-to-end execution.} Can OpenMQPI successfully compile and execute quantum applications through the OpenMP-based offloading pipeline on a real quantum resource?
\textbf{RQ2: Runtime overhead and performance.} What runtime overhead does OpenMQPI introduce compared with established quantum programming frameworks?
To answer these questions, we first validate the complete compilation and execution pipeline on a real quantum computing environment.
We then compare the runtime performance of OpenMQPI with Qiskit and CUDA Quantum across three representative quantum applications: \gls{vqe} , \gls{qft}, and Deutsch--Jozsa.

\subsection{Experimental Setup}
All experiments were conducted in a Docker environment running Ubuntu 22.04 on a MacBook Pro equipped with an Apple M3 Pro processor and 18 GB of memory.
OpenMQPI was built using LLVM 22.1.8, including \gls{mlir} and the OpenMP runtime infrastructure, together with MQSS Client interface.
%
For the validation experiment, quantum circuits were executed on a 54-qubit quantum resource.

For the performance experiments, we used a simulator to avoid variability due to queue waiting times for quantum resources.
We compared OpenMQPI with Qiskit 1.3.3 and CUDA Quantum 0.15.0.
All three implementations submitted their quantum circuits to the same simulator, providing a common execution environment for the comparison.

The evaluation covers both hybrid and pure quantum applications:
\gls{vqe} represents a hybrid quantum-classical application, whereas \gls{qft} and Deutsch--Jozsa are pure quantum applications.
All three applications were evaluated using 5, 10, 15, and 20 qubits.

\subsubsection{VQE Setup}
For \gls{vqe}  experiments, all implementations used the same parameterized ansatz, shown in Fig.~\ref{fig:vqe-ansatz}.
The ansatz consists of two parameterized $R_y$ layers separated by a linear chain of controlled-X (CX) gates between neighboring qubits.
For an $n$-qubit circuit, the ansatz contains $2n$ trainable parameters, all initialized to 0.25.
\begin{center}
\includesvg[width=0.90\columnwidth]{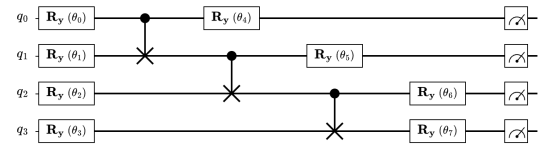}
    \captionof{figure}{Parameterized two-layer $R_Y$ ansatz used in the VQE evaluation for a four-qubit example.}
    \label{fig:vqe-ansatz}
\end{center}

\gls{vqe} objective function was defined using the following Hamiltonian:
\begin{equation}
    \hat{H} =
    -\sum_{i=0}^{n-1} Z_i
    -\frac{1}{2}\sum_{i=0}^{n-2} Z_i Z_{i+1}.
\end{equation}
For each parameter set, the quantum circuit was executed using 1,000 shots.
The resulting measurement counts were used to calculate the expectation values of the $Z_i$ and $Z_iZ_{i+1}$ terms, from which the corresponding energy was computed.
Since the Hamiltonian contains only $Z$ and $ZZ$ terms, these expectation values were obtained directly from the measured bitstrings without additional basis-rotation circuits.

All implementations used the COBYLA optimization algorithm.
OpenMQPI used the COBYLA implementation from the NLopt library, while Qiskit and CUDA Quantum used the COBYLA optimizers provided by their respective software stacks.
For all implementations, COBYLA was configured with a maximum of 100 objective-function evaluations, an initial step size of 0.25, and a relative parameter tolerance of $10^{-4}$.
\subsubsection{QFT and Deutsch--Jozsa Setup}

For \gls{qft} experiments, all implementations used the same circuit structure, shown in Fig.~\ref{fig:qft-circuit} for a four-qubit example.
The circuit decomposes the \gls{qft} into Hadamard, controlled-phase (CP), and SWAP gates.
In the figure, the black node denotes the control qubit, while the blue node denotes the target qubit to which the phase rotation is applied.
The rotation angle of each CP gate depends on the distance between the interacting qubits and is defined as
\[
\theta_{i,j} = \frac{\pi}{2^{j-i}}.
\]
After the Hadamard and controlled-phase operations, SWAP gates reverse the order of the output qubits, and all qubits are measured.
\begin{center}
\includesvg[width=0.90\columnwidth]{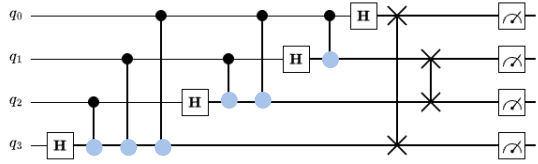}
    \captionof{figure}{QFT circuit used in the evaluation for a four-qubit example.}
    \label{fig:qft-circuit}
\end{center}

For the Deutsch--Jozsa experiments, all implementations used the same circuit structure, shown in Fig.~\ref{fig:dj-circuit} for a four-qubit example.
The circuit uses three input qubits and one ancilla qubit.
The ancilla qubit is first initialized to $|1\rangle$, after which Hadamard gates are applied to all qubits.
We used a balanced oracle, implemented using controlled-X (CX) gates between each input qubit and the ancilla qubit.
After the oracle, a second Hadamard gate is applied to each input qubit.
Finally, only the input qubits are measured, while the ancilla qubit is left unmeasured.
For both \gls{qft} and Deutsch--Jozsa, each circuit was executed using 1,000 shots.

\begin{center}
\includesvg[width=0.90\columnwidth]{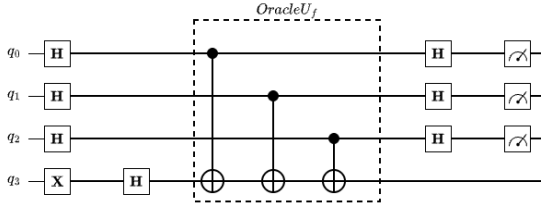}
    \captionof{figure}{Deutsch--Jozsa circuit used in the evaluation for a four-qubit example.}
    \label{fig:dj-circuit}
\end{center}

\subsection{Validation Experiment}

The validation experiment examines the end-to-end execution of OpenMQPI on a real quantum resource.
To analyze the runtime behavior of the pipeline, we instrumented the MQSS plugin and the host application to measure the execution time of its main stages.
For the validation experiments, we collected the following runtime metrics:
\begin{itemize}
    \item \textbf{Total Execution (TE):} End-to-end execution time of the application.
    
    \item \textbf{Lowering (TL):} Time required to lower the QPIR circuit to OpenQASM and prepare it for offloading.
    
    \item \textbf{Submit and Wait (TSW):} Time between submitting the quantum circuit and receiving the execution results.
    
    \item \textbf{Serialization (TS):} Time required to process the returned measurement results into the format expected by the application.
    
    \item \textbf{Constant Time (TC):} Remaining execution overhead not included in lowering, quantum job submission, or energy estimation.
    
    \item \textbf{Energy Estimation (TEE):} Time required to estimate the energy from the measurement results. This metric applies only to VQE.
\end{itemize}
\begin{table}[!ht]
    \centering
    \caption{Runtime metrics collected during the VQE validation experiment for different qubit counts.}
    \label{table:vqe-validation}

    \scriptsize
    \setlength{\tabcolsep}{2pt}

    \resizebox{\columnwidth}{!}{%
    \begin{tabular}{lcccccc}
        \toprule
        \textbf{Algorithms}
        & \textbf{TE (s)}
        & \textbf{TL (s)}
        & \textbf{TSW (s)}
        & \textbf{TS (s)}
        & \textbf{TC (s)}
        & \textbf{TEE (s)} \\
        \midrule

        VQE (5 qubits)
        & 124.41
        & $4.00\times10^{-3}$
        & 124.40
        & $5.96\times10^{-4}$
        & $1.38\times10^{-2}$
        & $2.00\times10^{-5}$ \\

        VQE (10 qubits)
        & 109.01
        & $3.54\times10^{-3}$
        & 108.99
        & $8.55\times10^{-4}$
        & $1.08\times10^{-2}$
        & $1.14\times10^{-4}$ \\

        VQE (15 qubits)
        & 76.20
        & $3.68\times10^{-3}$
        & 76.19
        & $1.01\times10^{-3}$
        & $1.13\times10^{-2}$
        & $3.80\times10^{-4}$ \\

        VQE (20 qubits)
        & 78.68
        & $3.73\times10^{-3}$
        & 78.66
        & $1.12\times10^{-3}$
        & $1.27\times10^{-2}$
        & $5.52\times10^{-4}$ \\

        \bottomrule
    \end{tabular}%
    }
\end{table}
Table~\ref{table:vqe-validation} presents the runtime metrics collected for the \gls{vqe} validation experiments with 5, 10, 15, and 20 qubits.
Across all configurations, the submit-and-wait time (TSW) accounts for most of the total execution time, while lowering, serialization, energy estimation, and the remaining execution overhead contribute only a small fraction of the overall runtime.
The variation in total execution time across qubit configurations is therefore primarily due to time spent waiting for the quantum resource.

\begin{table}[!ht]
    \centering
    \caption{Runtime metrics collected during the QFT and Deutsch--Jozsa
    validation experiments for different qubit counts.}
    \label{table:qft_dj_validation}

    \scriptsize
    \setlength{\tabcolsep}{2pt}

    \resizebox{\columnwidth}{!}{%
    \begin{tabular}{lccccc}
        \toprule
        \textbf{Algorithms}
        & \textbf{TE (s)}
        & \textbf{TL (s)}
        & \textbf{TSW (s)}
        & \textbf{TS (s)}
        & \textbf{TC (s)} \\
        \midrule

        QFT (5 qubits)
        & 6.10
        & $2.02\times10^{-3}$
        & 6.09
        & $3.71\times10^{-4}$
        & $8.83\times10^{-3}$ \\

        QFT (10 qubits)
        & 161.47
        & $2.33\times10^{-3}$
        & 161.46
        & $4.13\times10^{-4}$
        & $9.54\times10^{-3}$ \\

        QFT (15 qubits)
        & 86.25
        & $2.71\times10^{-3}$
        & 86.24
        & $6.01\times10^{-4}$
        & $7.15\times10^{-3}$ \\

        QFT (20 qubits)
        & 6.24
        & $2.77\times10^{-3}$
        & 6.23
        & $5.76\times10^{-4}$
        & $7.14\times10^{-3}$ \\

        \midrule

        Deutsch--Jozsa (5 qubits)
        & 877.63
        & $1.68\times10^{-3}$
        & 877.62
        & $3.84\times10^{-4}$
        & $8.24\times10^{-3}$ \\

        Deutsch--Jozsa (10 qubits)
        & 973.55
        & $1.68\times10^{-3}$
        & 973.53
        & $2.85\times10^{-4}$
        & $6.42\times10^{-3}$ \\

        Deutsch--Jozsa (15 qubits)
        & 1002.72
        & $1.99\times10^{-3}$
        & 1002.72
        & $3.42\times10^{-4}$
        & $6.90\times10^{-3}$ \\

        Deutsch--Jozsa (20 qubits)
        & 29.58
        & $1.78\times10^{-3}$
        & 29.58
        & $3.17\times10^{-4}$
        & $6.69\times10^{-3}$ \\

        \bottomrule
    \end{tabular}%
    }
\end{table}
Similar to \gls{vqe}, Table~\ref{table:qft_dj_validation} presents the runtime metrics collected for the \gls{qft} and Deutsch--Jozsa validation experiments, and the TSW accounts for most of the total execution time across all configurations, while lowering, serialization, and the remaining execution overhead contribute only a small fraction of the overall runtime.
Since \gls{qft} and Deutsch--Jozsa do not perform energy estimation, the TEE metric is not applicable to these applications.

Overall, the successful execution of all three applications on a quantum resource validates the complete OpenMQPI compilation and execution pipeline.

\subsection{Performance Evaluation}
To characterize runtime performance, each application and qubit configuration was executed 100 times, and the execution time of each run was recorded for the analysis.

\begin{center}
    \includegraphics[width=\columnwidth]{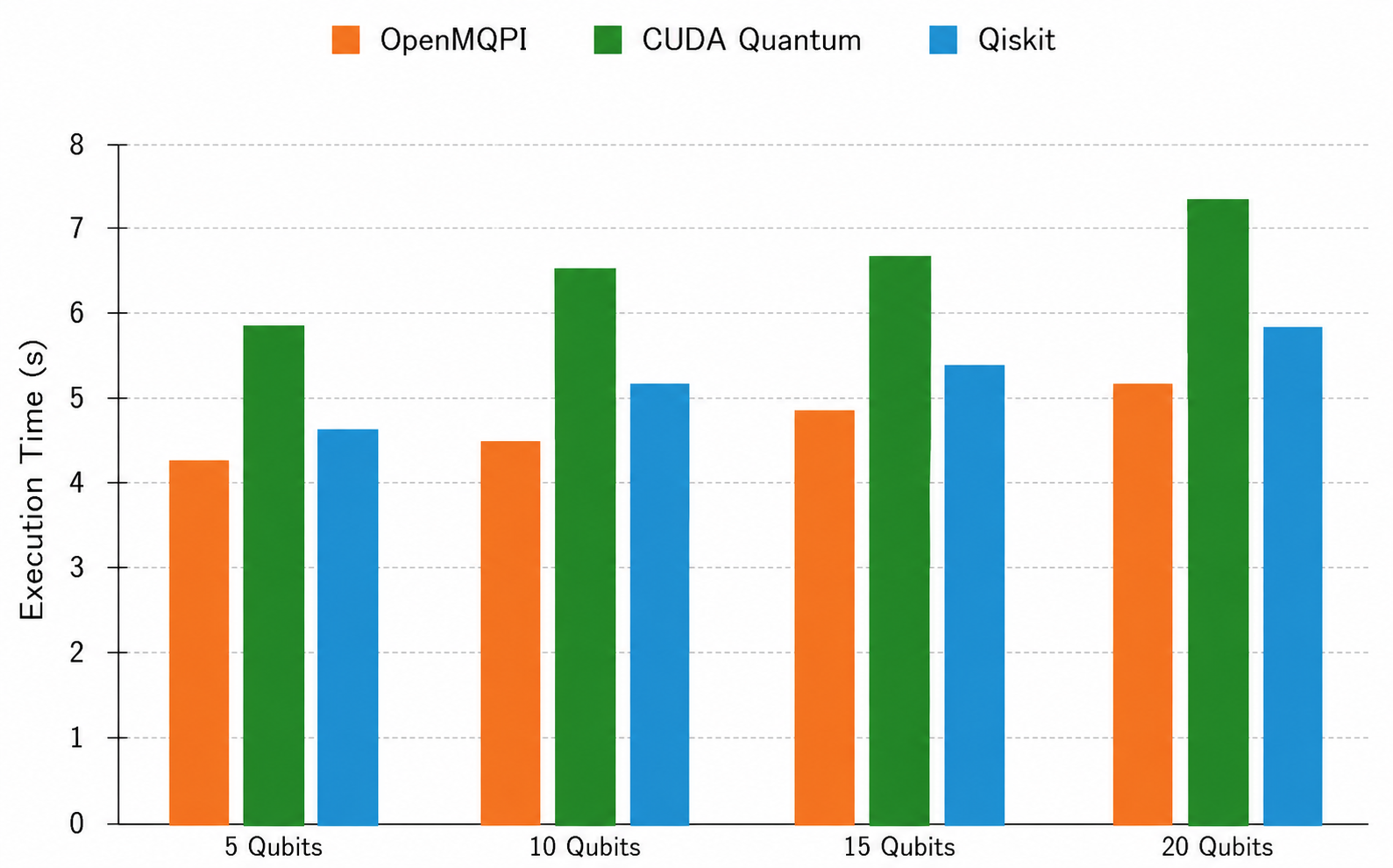}
    \captionof{figure}{Average VQE execution time for different qubit configurations.}
    \label{fig:vqe-runtime}
\end{center}

Fig.~\ref{fig:vqe-runtime} compares the average \gls{vqe} execution times of OpenMQPI, Qiskit, and CUDA Quantum for 5, 10, 15, and 20 qubits.
%
Under the evaluated experimental setup, OpenMQPI achieves the lowest average execution time across the evaluated qubit configurations.
Its execution time increases from 4.37~s for 5 qubits to 5.18~s for 20 qubits.
Qiskit ranges from 4.67~s to 5.83~s, while CUDA Quantum ranges from 5.86~s to 7.39~s.

\begin{figure*}[h]
    \centering
    \begin{minipage}[t]{0.48\textwidth}
        \centering
        \includegraphics[width=\linewidth]{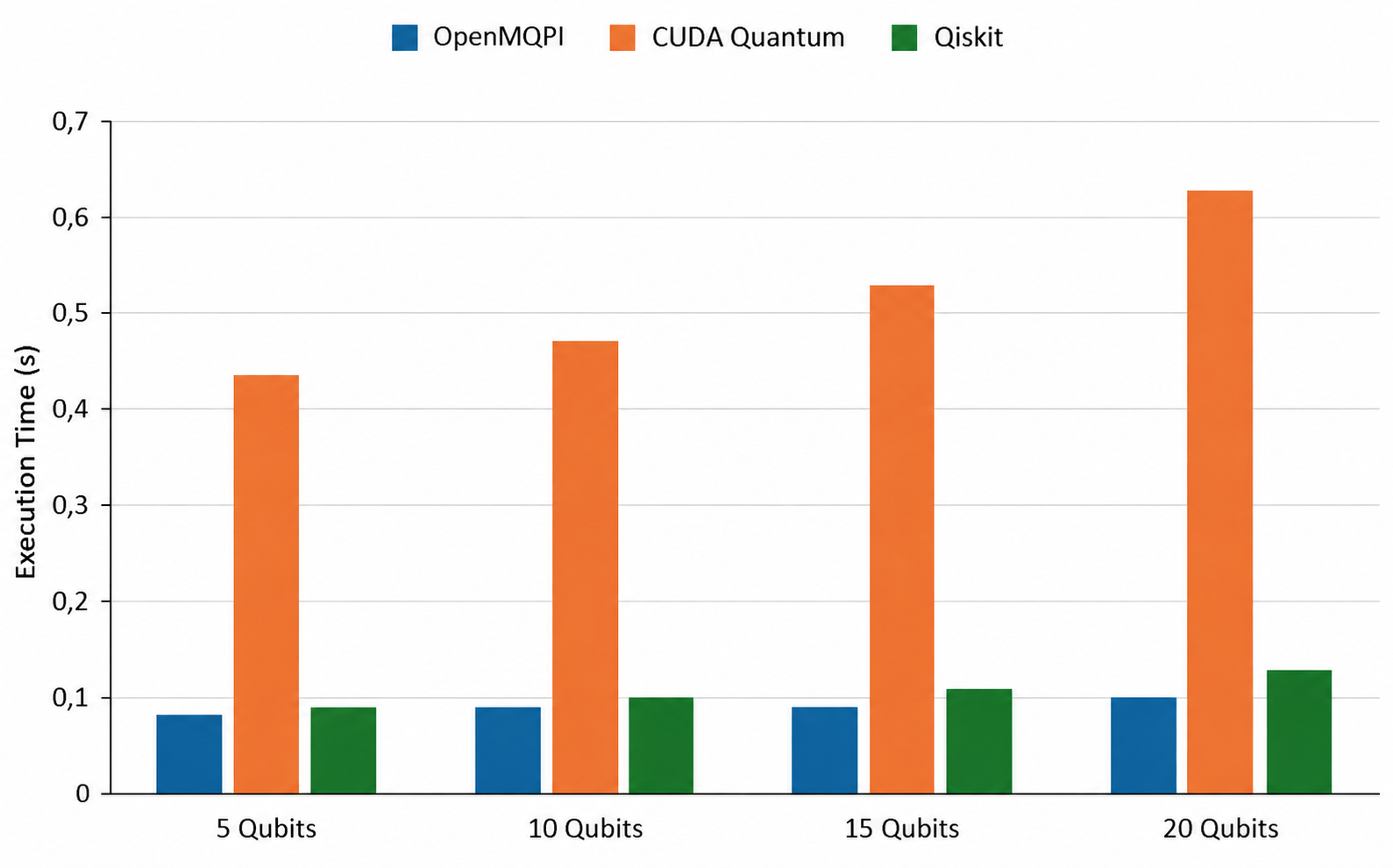}
        \caption{Average QFT execution time for different qubit configurations.}
        \label{fig:qft-runtime}
    \end{minipage}
    \hfill
    \begin{minipage}[t]{0.48\textwidth}
        \centering
        \includegraphics[width=\linewidth]{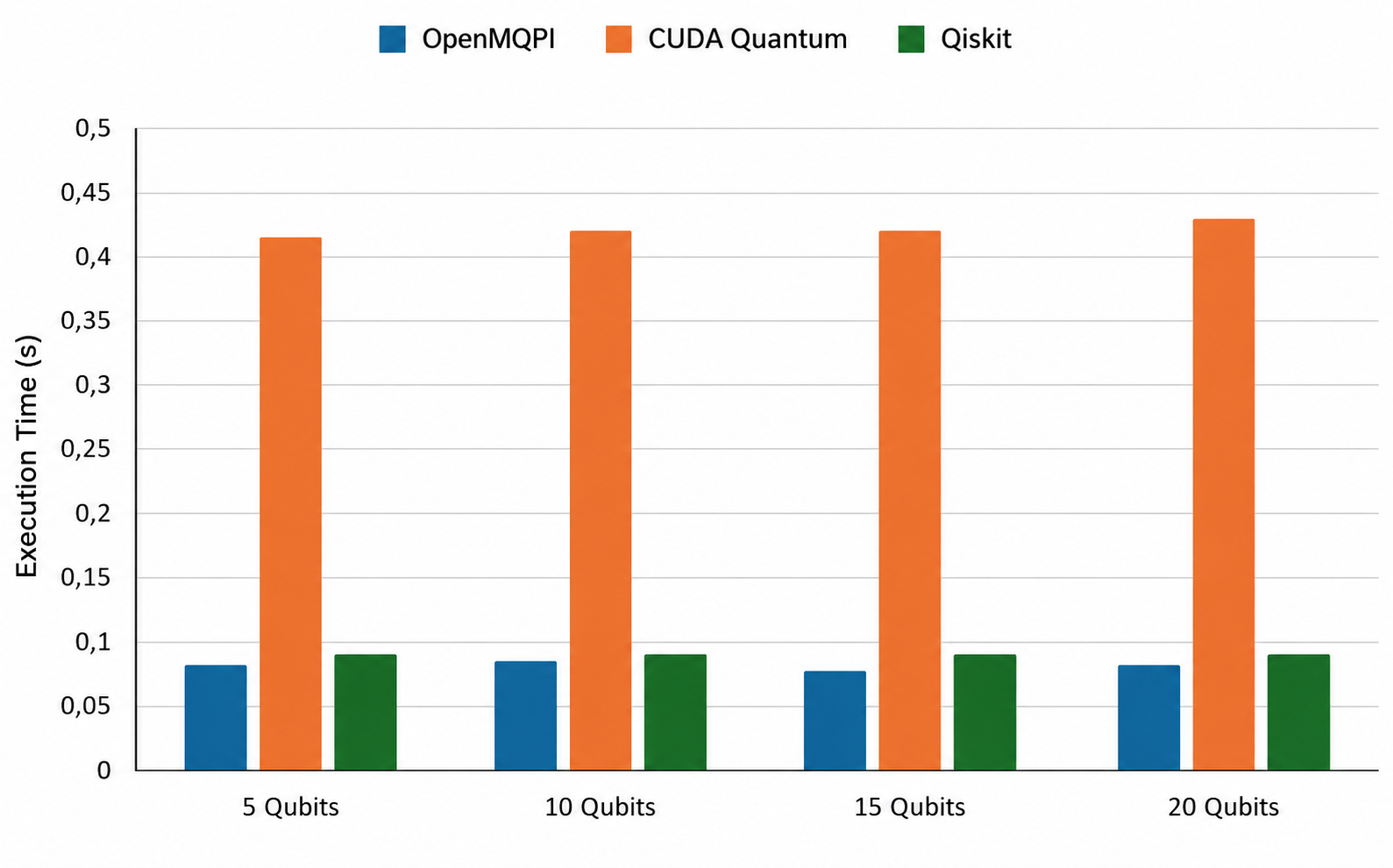}
        \caption{Average Deutsch–Jozsa execution time for different qubit configurations.}
        \label{fig:dj-runtime}
    \end{minipage}

\end{figure*}
\begin{table}[!ht]
    \centering
    \caption{Average runtime breakdown of the 5-qubit VQE benchmark for the evaluated implementations.}
    \label{table:vqe_metrics}

    \small
    \setlength{\tabcolsep}{2.5pt}

    \resizebox{\columnwidth}{!}{%
    \begin{tabular}{lcccccc}
        \toprule
        & \multicolumn{6}{c}{\textbf{Metrics}} \\
        \cmidrule(lr){2-7}

        \textbf{Implementation}
        & \textbf{TE (s)}
        & \textbf{TL (s)}
        & \textbf{TSW (s)}
        & \textbf{TS (s)}
        & \textbf{TC (s)}
        & \textbf{TEE (s)} \\
        \midrule

        OpenMQPI
        & 4.35
        & $7.30\times10^{-3}$
        & 4.25
        & $1.28\times10^{-3}$
        & $9.33\times10^{-2}$
        & $1.45\times10^{-4}$ \\

        CUDA Quantum
        & 5.79
        & 1.71
        & 3.97
        & $7.62\times10^{-3}$
        & $9.38\times10^{-2}$
        & $1.69\times10^{-3}$ \\

        Qiskit
        & 5.00
        & $7.89\times10^{-2}$
        & 4.34
        & --
        & $9.11\times10^{-2}$
        & $1.76\times10^{-2}$ \\

        \bottomrule
    \end{tabular}%
    }
\end{table}
\begin{table*}[t]
    \centering
    \caption{Average runtime breakdown of the 5-qubit QFT and Deutsch--Jozsa benchmarks.}
    \label{tab:qft_dj_metrics}

    \small
    \setlength{\tabcolsep}{8pt}

    \begin{tabular}{llccccc}
        \toprule
        \textbf{Algorithm}
        & \textbf{Implementation}
        & \multicolumn{5}{c}{\textbf{Metrics}} \\
        \cmidrule(lr){3-7}

        &
        & \textbf{TE (s)}
        & \textbf{TL (s)}
        & \textbf{TSW (s)}
        & \textbf{TS (s)}
        & \textbf{TC (s)} \\
        \midrule

        \multirow{3}{*}{QFT}
        & OpenMQPI
        & $7.79\times10^{-2}$
        & $1.93\times10^{-3}$
        & $6.76\times10^{-2}$
        & $1.54\times10^{-4}$
        & $8.22\times10^{-3}$ \\

        & CUDA Quantum
        & $4.22\times10^{-1}$
        & $3.62\times10^{-1}$
        & $4.51\times10^{-2}$
        & $1.87\times10^{-3}$
        & $1.30\times10^{-2}$ \\

        & Qiskit
        & $8.40\times10^{-2}$
        & $4.53\times10^{-3}$
        & $4.55\times10^{-2}$
        & --
        & $3.40\times10^{-2}$ \\

        \midrule

        \multirow{3}{*}{Deutsch--Jozsa}
        & OpenMQPI
        & $7.83\times10^{-2}$
        & $1.77\times10^{-3}$
        & $6.82\times10^{-2}$
        & $1.51\times10^{-4}$
        & $8.16\times10^{-3}$ \\

        & CUDA Quantum
        & $4.05\times10^{-1}$
        & $3.40\times10^{-1}$
        & $4.68\times10^{-2}$
        & $1.90\times10^{-3}$
        & $1.63\times10^{-2}$ \\

        & Qiskit
        & $8.52\times10^{-2}$
        & $3.56\times10^{-3}$
        & $4.56\times10^{-2}$
        & --
        & $3.51\times10^{-2}$ \\

        \bottomrule
    \end{tabular}
\end{table*}
Table~\ref{table:vqe_metrics} presents the average runtime breakdown for the 5-qubit VQE benchmark.
OpenMQPI has the lowest lowering time, requiring $7.30\times10^{-3}$~s compared with $7.89\times10^{-2}$~s for Qiskit and 1.71~s for CUDA Quantum.
OpenMQPI uses QPIR, a lightweight \gls{mlir} dialect designed specifically for the proposed compilation and offloading pipeline.
In comparison, CUDA Quantum represents quantum operations using its Quake dialect, while Qiskit represents circuits using its own circuit model and processes them through its transpilation infrastructure.
Qiskit does not expose a separate serialization stage in our measurements, as measurement result processing is handled within its execution pipeline.
In contrast, OpenMQPI and CUDA Quantum perform serialization as a separate step.
OpenMQPI requires $1.28\times10^{-3}$~s for this stage, compared with $7.62\times10^{-3}$~s for CUDA Quantum.
Overall, the \gls{vqe} results show that OpenMQPI introduces low compilation and runtime processing overhead and achieves lower end-to-end execution times than Qiskit and CUDA Quantum across the evaluated qubit configurations.

Figures~\ref{fig:qft-runtime} and~\ref{fig:dj-runtime} present the average execution times of QFT and Deutsch--Jozsa, respectively, for 5, 10, 15, and 20 qubits.

For both \gls{qft} and Deutsch--Jozsa, OpenMQPI shows lower average execution times than CUDA Quantum across all evaluated qubit configurations, while
its execution times remain close to those of Qiskit.
For \gls{qft}, the execution time of CUDA Quantum increases more noticeably with the number of qubits, whereas OpenMQPI and Qiskit show smaller increases.
For Deutsch--Jozsa, the execution times remain relatively stable across the evaluated qubit configurations for all three implementations.

To further examine these differences, Table~\ref{tab:qft_dj_metrics} presents the average runtime breakdown for the 5-qubit \gls{qft} and Deutsch--Jozsa benchmarks.
For both applications, OpenMQPI shows the lowest lowering time among the evaluated implementations.
For \gls{qft}, the lowering time is $1.93\times10^{-3}$~s for OpenMQPI, compared with $4.53\times10^{-3}$~s for Qiskit and $3.62\times10^{-1}$~s for CUDA Quantum.
A similar pattern is observed for Deutsch--Jozsa, where OpenMQPI requires $1.77\times10^{-3}$~s, compared with $3.56\times10^{-3}$~s for Qiskit and $3.40\times10^{-1}$~s for CUDA Quantum.
This is consistent with the \gls{vqe} results, where OpenMQPI also shows a relatively small lowering overhead.
As in the \gls{vqe} benchmark, Qiskit does not expose a separate serialization time, whereas OpenMQPI and CUDA Quantum perform serialization as a separate stage.
For both \gls{qft} and Deutsch--Jozsa, the serialization overhead of OpenMQPI remains small.
The submit-and-wait times are of similar magnitude across the three implementations, which use the same simulator as the execution environment.
The remaining constant-time overhead is also small for OpenMQPI.
Unlike \gls{vqe}, \gls{qft}, and Deutsch--Jozsa do not perform energy estimation; therefore, the TEE metric is not applicable to these benchmarks.

\section{Related Works}
\label{sec:related-works}

Several quantum software frameworks and interfaces have been developed to support the programming and execution of quantum applications across different quantum computing environments.
These frameworks and interfaces provide different abstractions for representing quantum programs, transforming quantum circuits, and interacting with quantum computing resources.

One of these frameworks is Qiskit, an open-source software development kit for developing and executing quantum applications.
It provides a circuit-based programming model in which quantum computations are represented as circuits composed of quantum and classical operations ~\citep{qiskit}.
Qiskit includes a transpiler that transforms and optimizes quantum circuits according to the instruction set and constraints of a target quantum device.
While Qiskit performs these transformations within its circuit transpilation infrastructure, OpenMQPI represents and optimizes quantum operations directly within \gls{mlir} using the \gls{qpir} dialect.
Qiskit also supports execution on both quantum simulators and physical quantum hardware, as well as hybrid quantum-classical workflows that combine quantum circuit execution with classical computation ~\citep{pathak2025evolutionibmsquantuminformation}.
Its modular design also allows the framework to be extended with new circuit synthesis and optimization methods and to target different quantum hardware backends.

\begin{figure}[!ht]
    \centering
    \includegraphics[width=0.55\columnwidth]{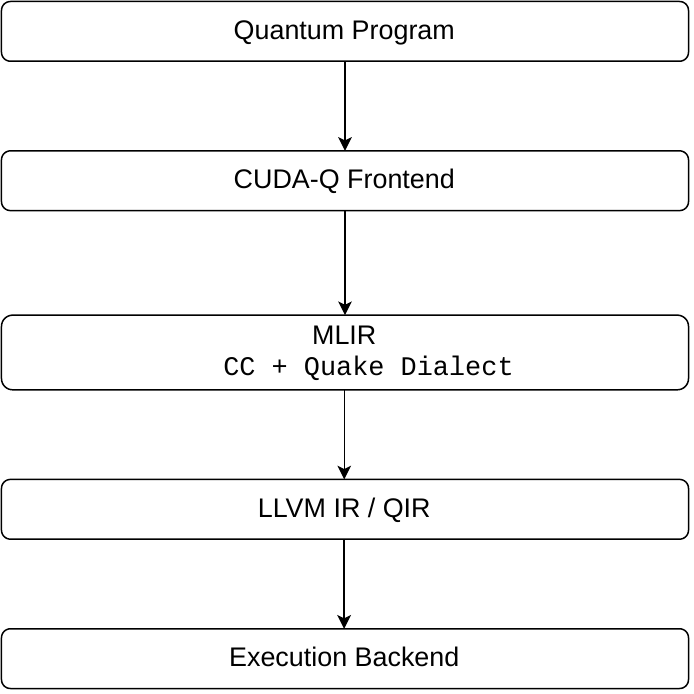}
    \caption{Simplified overview of the CUDA-Q compilation workflow.}   \label{fig:cuda_quantum_flow}
\end{figure}

Another framework is CUDA Quantum, a programming model and software platform designed for heterogeneous quantum-classical computing \citep{cuda-q}.
CUDA Quantum integrates quantum resources into \gls{hpc} and \gls{gpu}-accelerated workflows, providing interoperability with classical parallel programming models.
At the compiler level, CUDA Quantum represents quantum programs using \gls{mlir} \citep{lücke2024mlirtransformdialectcompiler} and the Quake \citep{nvidia_quake_dialect} dialect for quantum operations \citep{kulkarni2026efficienttranspilationopenqasm30}, as illustrated in Fig. ~\ref{fig:cuda_quantum_flow}.
The resulting representation is then lowered through LLVM and emitted as \gls{qir} for execution on simulation or quantum hardware backends.

%
%
In HPC environments, the \gls{qpi}~\citep{10821019} offers a C-based runtime interface for integrating quantum programs into \gls{hpc} applications.
Similarly, CQ~\citep{cq} provides a C-like API, programming model, and specification for developing hybrid applications.


\gls{mqss-db} is a modular and open-source software ecosystem designed to support hybrid quantum-classical applications and integrate quantum computing resources into \gls{hpc} environments \citep{Burgholzer_2026, disaggregated-paper}.
%
%
%
Its architecture decouples high-level programming interfaces from the underlying quantum hardware while supporting compilation, resource management, and scheduling within the same software stack.

Similarly, \gls{qrmi} \citep{bacher2025quantum} provides a vendor-agnostic library to access, control, and monitor the quantum resources.\
It is designed to hide the vendor-specific implementation details by providing interfaces for \gls{hpc} environments.

%
%
%


Mittal et al. propose OpenMP-Q, which extends the OpenMP target model to support the offloading of quantum tasks from C/C++ applications to quantum devices \citep{mittal2025openmp}.
It introduces a \texttt{circuit} clause to the OpenMP \texttt{target} directive, which passes a user-defined quantum circuit to the OpenMP runtime for offloading.
At runtime, OpenMP-Q uses a POSIX pipe-based bridge to connect the C/C++ OpenMP application to Python-based quantum software for circuit execution.
OpenMP-Q also supports reverse offloading for iterative quantum-classical workflows and the parallel execution of quantum tasks across multiple QPUs.

%
Lee et al. propose Quantum Task Offloading with the OpenMP API to introduce an OpenMP-based programming model to integrate quantum computing resources into \gls{hpc} applications \citep{lee2023quantumtaskoffloadingopenmp}.
The approach uses the OpenMP target construct to treat quantum devices as offloading targets, following the same host-device execution model used for conventional accelerators such as \glspl{gpu}.
Within the target region, quantum circuits are constructed via dedicated function calls to create quantum registers, apply quantum gates, and measure.
These quantum operations are then transpiled into OpenQASM and \gls{qir}, which can be passed to their quantum resource manager for further optimization and scheduling on a quantum device or simulator.
This programming model is closely related to OpenMQPI, as both approaches integrate quantum execution into the OpenMP target-offloading model.
However, while this approach represents quantum operations via dedicated function calls that are subsequently transpiled to QASM or \gls{qir}, OpenMQPI represents them directly as QPIR operations within \gls{mlir}.

OpenMQPI differs from these approaches by introducing QPIR as an MLIR-level representation of quantum operations and by preserving this representation through compiler lowering until the circuit is serialized for the OpenMP runtime.

\section{Conclusion and Future Work}
\label{sec:conclusion}

Integrating quantum computing resources into \gls{hpc} applications remains challenging for portability and reusability, as existing quantum programming frameworks rely on different programming models, compiler infrastructures, and execution environments.
OpenMP is a well-established programming model in \gls{hpc} that provides a portable and reusable approach to parallel programming and target offloading.
Building on this existing infrastructure, this paper presented OpenMQPI, an \gls{mlir}-based compilation infrastructure that integrates quantum circuit compilation and execution into the OpenMP target-offloading model.

OpenMQPI uses the \gls{qpir} dialect to represent quantum circuits directly within \gls{mlir}, enabling quantum-specific optimizations before the circuits are lowered for execution.
The resulting circuits are integrated into the OpenMP offloading pipeline and executed on quantum resources through \texttt{libomptarget} and MQSS Client interface.
%
The evaluation demonstrated that OpenMQPI successfully executes \gls{vqe}, \gls{qft}, and Deutsch--Jozsa through the proposed compilation and offloading pipeline on a real quantum computing environment.
%
Under the evaluated simulator-based configuration, OpenMQPI achieves competitive runtime performance compared with Qiskit and CUDA Quantum.
Overall, OpenMQPI demonstrates that quantum computing resources can be integrated into the established OpenMP offloading infrastructure while preserving a compiler-based workflow for quantum circuit representation, optimization, and execution.

Future work will focus on extending OpenMQPI in several directions.
First, integrating the OpenMQPI front end more directly with Clang could provide a more unified compilation workflow for quantum-enabled OpenMP applications.
Second, the lowering pipeline could be extended to support \gls{qir} as an additional representation, enabling interoperability with a broader range of quantum software and execution environments.
Finally, integrating a local quantum simulator would allow quantum circuits to be executed directly within the OpenMQPI environment without requiring access to an external quantum resource.
These extensions would further strengthen OpenMQPI as a compiler-based approach for integrating quantum computing resources into established HPC programming and offloading environments.

\begin{acks}
This work is supported by the German Federal Ministry of Research, Technology, and Space (BMFTR) with the grants 13N15689 (DAQC), 13N16063 (Q-Exa), 13N16078 (MUNIQC-Atoms), 13N16187 (MUNIQC-SC), 13N16690 (Euro-Q-Exa), 13N16894 (MAQCS), European fundings 101114305 (Millenion), 10111394\-6 (OpenSuperQPlus), 101194491 (QEX), and the Bavarian State Ministry of Science and the Arts (StMWK) through funding, as part of Munich Quantum Valley, Q-DESSI.
\end{acks}

\bibliographystyle{ACM-Reference-Format}
\bibliography{references}

\end{document}